\documentclass[10pt,preprintnumbers,showpacs,amsmath,amssymb,floatfix,twocolumn,prl]{revtex4}
\newcommand{\fig}[1]{Fig. \ref{fig:#1}}

\newcommand{\ie}{\textit{i.e. }}

\newcommand{\expect}[1]{\langle #1 \rangle}

\newcommand{\beq}{\begin{equation}}
\newcommand{\eeq}{\end{equation}}

\usepackage{graphicx}
\usepackage{epstopdf}

\begin{document}
\title{Models of organometallic complexes for optoelectronic applications}
\author{A. C. Jacko}
\affiliation{Centre for Organic Photonics and Electronics, School of Mathematics and Physics, The University of Queensland}
\author{Ross H. McKenzie}
\affiliation{Centre for Organic Photonics and Electronics, School of Mathematics and Physics, The University of Queensland}
\author{B. J. Powell}
\email{bjpowell@gmail.com}
\affiliation{Centre for Organic Photonics and Electronics, School of Mathematics and Physics, The University of Queensland}
%\date{\today}

\begin{abstract}
Organometallic complexes have potential applications as the optically active components of organic light emitting diodes (OLEDs) and organic photovoltaics (OPV). Development of more effective complexes may be aided by understanding their excited state properties. Here we discuss two key theoretical approaches to investigate these complexes: first principles atomistic models and effective Hamiltonian models. We review applications of these methods, such as, determining the nature of the emitting state, predicting the fraction of injected charges that form triplet excitations, and explaining the sensitivity of device performance to small changes in the molecular structure of the organometallic complexes.

\end{abstract}

\maketitle

\section{Introduction}

The development of organic light emitting diodes\cite{tang87} and dye-sensitized solar cells\cite{gratzel91} two decades ago has lead to an explosion of interest in the optoelectronic properties of organic materials for use in photovoltaic and electroluminescent applications.\cite{hagfeldt00,li05,lo06} For example, a material with the right set of optoelectronic properties could revolutionize solar energy production, replacing fragile, high-temperature processed silicon with durable, solution processed, ink-jet printable plastic electronics.\cite{hebner98,friend99,forrest00,forrest04} However, the efficiency of optoelectronic devices needs to increase in order for them to become commercially viable beyond niche applications. Improving their efficiency while retaining the desired chemical and physical properties has proven challenging.\cite{forrest04,burn07}

\begin{figure}
	\centering
		\includegraphics[width=6.8cm]{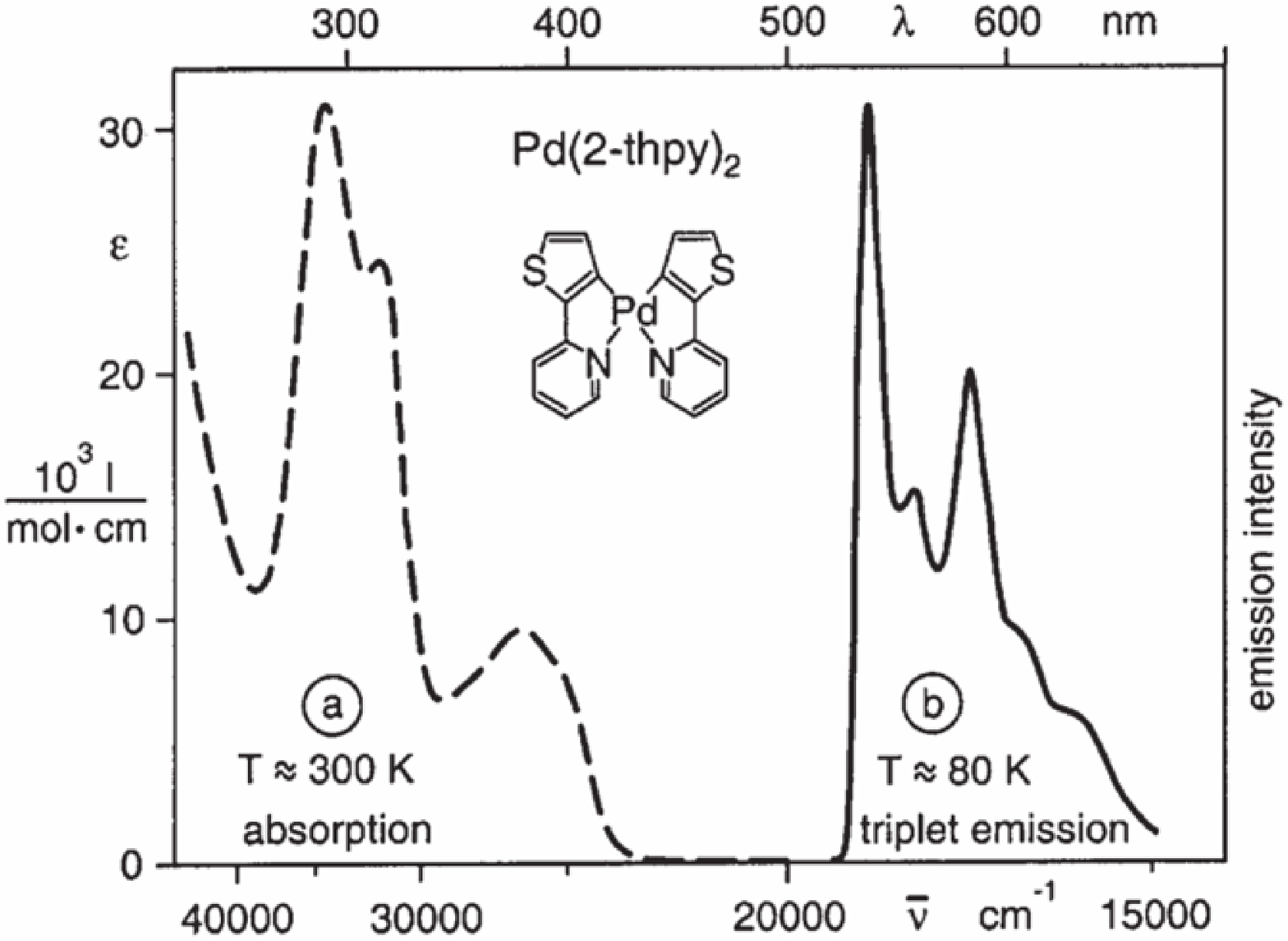}\\
		\includegraphics[width=6.4cm]{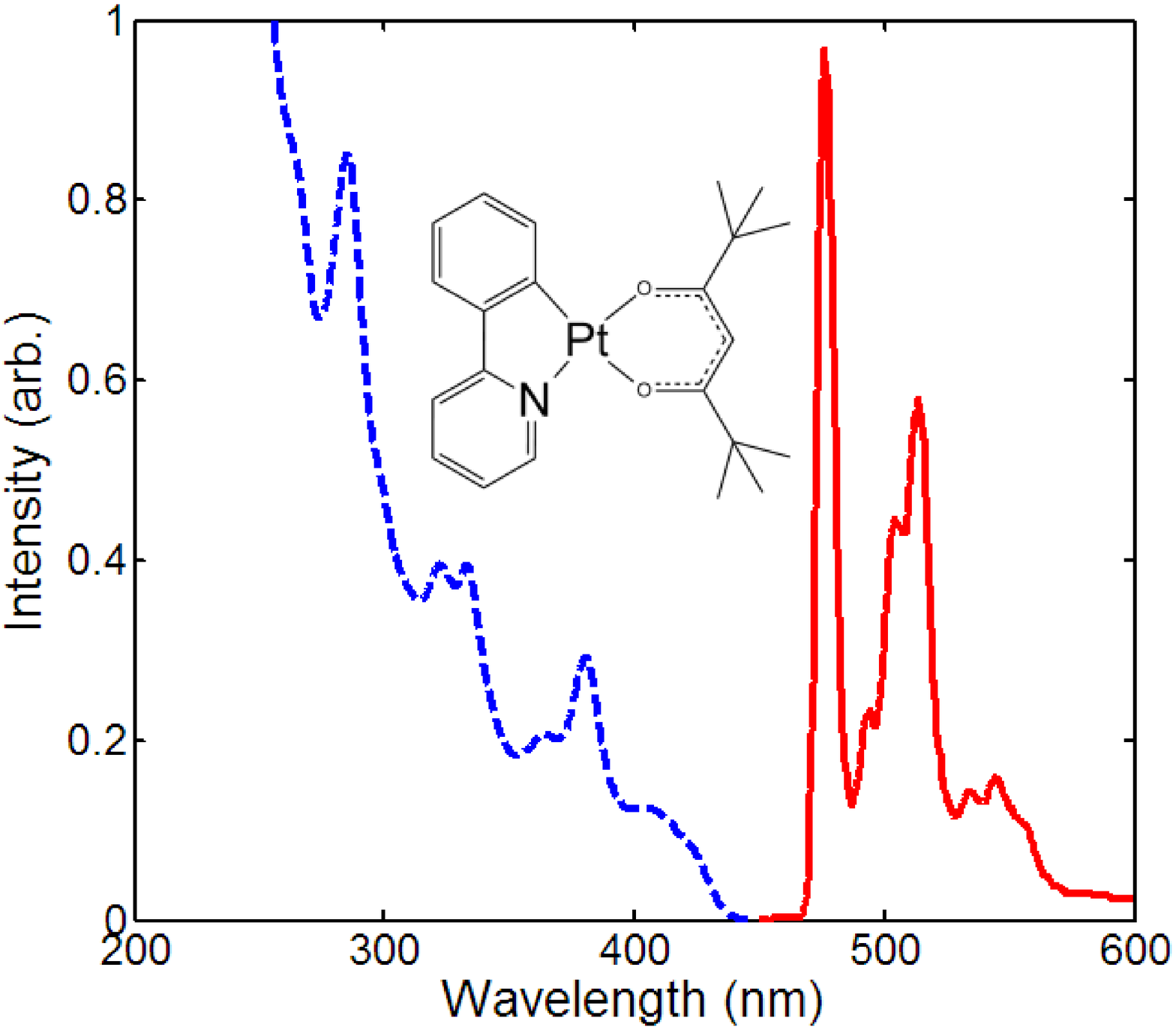}
			\caption{Absorption and emission spectra of Pd(2-thpy)$_2$(top panel) from Ref. \onlinecite{yersin01}, and ppyPt(dpm) (bottom panel) from Ref. \onlinecite{brooks02}. In both panels the dashed line shows the absorption spectra and the solid line shows emission spectra. The brighter higher energy absorption bands are usually identified as LC ($\pi\rightarrow \pi^*$) singlets, the less intense lower energy band is usually identified as MLCT singlets. The emission is typically considered to be predominantly triplet, and, depending on the complex may be LC, MLCT or a hybrid of the two.}
	\label{fig:complexes}
\end{figure}

Some of the most promising candidate materials for light-emitting devices are organometallic complexes. Fig. \ref{fig:complexes} shows the structures and spectra of two typical organometallic complexes used in OLEDs. The presence of a heavy transition metal ion core allows triplet states to radiatively decay (via a large spin-orbit coupling). It is widely supposed that upon injection triplets and singlet are formed in the ratio 3:1. Hence phosphorescence allows for a more efficient utilization of injected charges.\cite{sun06}

\begin{figure}
	\centering
		\includegraphics[width=6.3cm]{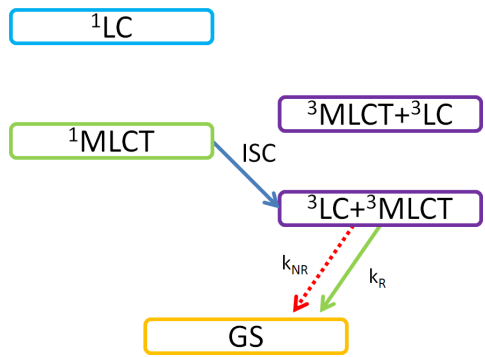}
	\caption{A schematic of the decay pathways in a phosphorescent material based on the results of Ref. \onlinecite{jacko10a}. Injected charges form either an MLCT singlet excited state ($^1 MLCT$) or a hybrid LC-MLCT triplet state ($^3 LC + ^3 MLCT$). MLCT singlets undergo a fast (fs) intersystem crossing (ISC) to the triplet state, the speed of which is determined by the magnitude of the spin-orbit coupling and the degree of MLCT character in the triplet state.\cite{yoon06,haneder08,jacko10a} The triplet excited state relaxes to the ground state (GS), either radiatively, with a rate $k_R$, or non-radiatively, with a rate $k_{NR}$. The photoluminescent quantum yield (PLQY) of the complex is then $\frac{k_R}{k_R + k_{NR}}$.}
	\label{fig:decaypath}
\end{figure}

The technologically useful properties of organometallic complexes, namely their absorption, emission and charge transfer properties, are determined by the excited states of the complex. For example, the process of electroluminescence involves the injection of positive and negative charges into a bulk sample of the complex, which results in the ionized complexes. These charges migrate until they combine on a single complex, leaving it in an electronically excited state. This excited complex then decays back to it's ground state by emitting a photon. A key design criterion is that the radiative decay must be much faster than any non-radiative decay processes. In OPV devices, incoming photons create excited states, \emph{i.e.}, an electron-hole pair. These charges must then be separated to perform useful work. Thus, in OPV devices, one wants charge-seperation to be the preferred decay pathway for excited states.
\begin{figure}
	\centering
		\includegraphics[width=4.5 cm]{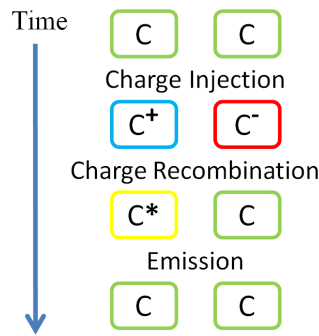}
	\caption{A schematic of the process associated with the operation of a LED. From top to bottom: The neutral complexes (C) are in their ground states. An electron is injected from the right, and a hole from the left. These ionic complexes are now in their new (charged) ground states. The electron and hole combine on a single complex (by either electron transfer between the LUMOs of adjacent complexes, or hole transfer between the HOMO-metal hybrid orbitals), forming a singlet or a triplet excited state (C*), while the other site returns to the neutral ground state. Finally, the excited state relaxes to the neutral ground state, either by emitting a photon or by non-radiative mechanisms.}
	\label{fig:chargeinjection}
\end{figure}
Fig. \ref{fig:chargeinjection} shows a schematic of charge recombination in an OLED. To understand and optimize this process (and the reverse process in OPVs), one must first understand the various excited states involved. Fig. \ref{fig:decaypath} shows a schematic of the likely decay pathway in typical phosphorescent OLED materials.

A key property of the excited states of organometallic complexes is their character, \ie the degree of singlet or triplet character and the degree of ligand centered (LC) or metal-to-ligand charge transfer (MLCT) character.  By understanding which parts of the complex influence each excited state, one might begin to engineer the spectrum of a complex by tuning the chemistry of the complex. Notice that there are two bands in both of the absorption spectra shown in Fig. \ref{fig:complexes}. The higher energy, more absorptive band is identified with LC ($\pi-\pi^*$) transitions. This higher energy (usually UV) band is present in spectra of isolated ligands, and thus its identification as a LC transition is uncontentious.  The lower energy, lower absorptivity band is usually associated with MLCT singlets as it is seen only in complexes.\cite{yersinbook} 
The classification of the emitting state has attracted considerable interest and debate, and it has been labeled variously as MLCT, LC or MLCT-LC hybrid.\cite{colombo93} 
The emitting state is believed to be a triplet, but the precise degree of triplet character has long been a question of interest.\cite{kober82}

The brightness of the MLCT singlet is due to the hybridization of (occupied) metal orbitals and (unoccupied) ligand $\pi^*$ (LUMO) orbital. This mixing delocalizes the excitation over the whole complex, allowing the excited state to have a transition to the ground state of similar brightness to a localized $\pi^* \rightarrow \pi$ (LC) transition. This shift of electron density from the metal to the ligand is known as $\pi$ back-bonding. This interaction has several experimental signatures, including an increase in the length of the C-C bond in the ligand and a softening of the ligand C-C stretch frequency as well as the bright MLCT transition.\cite{jean05,reimers91,zwickel71}

The presence of a heavy transition metal core in an organometallic complex produces a strong spin-orbit coupling, since the magnitude of the spin-orbit interaction increases proportional to $Z^4$ (where $Z$ is the atomic number of the element).\cite{yersinbook} The large spin-orbit interaction in the presence of heavy metal ion causes a hybridization of singlets and triplets.\cite{kober82} This has several important effects: First, the states dominated by singlets and triplets are now coupled by a rapid intersystem crossing ($\sim$50 fs) via their MLCT components.\cite{yoon06,yersinbook} Second, spin-orbit coupling allows triplet-dominated states to decay radiatively. This allows for phosphorescent OLEDs, increasing the excitation-to-photon conversion efficiency by harvesting triplet excitations.\cite{sun06}

% Yersin ZFM MLCT claim
Finkenzeller \emph{et al}.\cite{finkenzeller07} have argued that the degree of zero-field splitting in the emitting triplet reflects the amount of MLCT character in the state. They further state that there is `\emph{an empirical correlation between the amount of ZFS and the compound's potential for its use as emitter material in an OLED}'. They conclude that increasing the MLCT character of the emitting triplet is `a necessary but not sufficient condition' for improving a complexes suitability for use in OLED devices.\cite{finkenzeller07}

% clearly understanding the character of the excited states is vital
Understanding the character of the excited states may help make better materials for organic electronic devices. Furthermore, it may allow us to understand why empirical correlations, such as those of Finkenzeller \emph{et al}., are true and when we can expect them to be valid. Further, there remain many open questions about the properties of organometallic complexes including: 
\begin{itemize}
	\item Why are their emission properties so sensitive to small chemical changes, \emph{cf., e.g.}, Ref. \onlinecite{lo06}?
	\item What is the character of the emitting triplet state - to what extent is it MLCT, or LC? 
	\item What is the non-radiative decay path of the emitting state? What part do unoccupied non-bonding metal orbitals play in non-radiative decay? 
	\item What is the mechanism for ultrafast singlet-triplet MLCT transitions? Is there a conical intersection associated with this intersystem crossing? 
	\item Are the excited states localized over all the ligands, or localized on one? How does this change with perturbations to the symmetry of the complex? 
	\item What fraction of injected charges result in triplet excitations?
\end{itemize}
Clearly, theoretical input has an important role to play in answering these, and other, questions. To date, majority of modelling of organometallic complexes has been atomistic first principles quantum chemistry. However, another approach is to model Hamiltonians that capture the main features of broad classes of materials. In the remainder of this paper we compare and contrast how these two complimentary approaches have been used to address some of the above questions.

\section{First principles quantum chemistry}

First principles approaches have the advantage of being materials specific. This can be invaluable for understanding the role of chemical changes in the complexes. However, it is also a disadvantage in that it requires many different complexes to be studied before one can hope to access these trends, much like experiment.
The organometallic complexes used in OPV and OLED applications present a number of challenges to atomistic modelling. They are quite large, which limits the use of high quality \emph{ab initio} methods - although some calculations have been reported. \cite{pierloot03} Yet electronic correlations do play an important role, particularly in determining the excitation energies that determine the optical properties of the complex. Ideally one would like to be able to predict excitation energies to within chemical accuracy (\emph{i.e.}, within $\sim k_BT$).
By far the most widespread approaches to the first principles modelling of organometallic complexes have been density functional theory (DFT) for ground state properties and time-dependent density functional theory (TDDFT) for excited states properties (see, for example, Refs. \onlinecite{hay02,obara06,lo06,bomben09, wilson10, baccouche10, butschke10, mendes10, jacobsen10}). (TD)DFT gives reasonable accuracy at a low computational cost. While these methods do not provide chemical accuracy for the excited state properties they have allowed for trends across related complexes to be studied.
In particular, many researches have attempted to identify the character of excitations, \emph{i.e.}, whether they are ligand centered (LC), metal-to-ligand charge transfer (MLCT) or inter-ligand excitations, with (TD)DFT (see, for example, Refs. \onlinecite{hay02,obara06,lo06,bomben09, wilson10, baccouche10, butschke10, mendes10, jacobsen10}).

While TDDFT is a useful resource, current implementations have some well known shortcomings.\cite{jones89,ku02,ghosh06,wodrich06,cohen08,brittain09}  For example, it is well known that current (TD)DFT implementations do not correctly capture electronic correlations or describe charge transfer processes well.\cite{cohen08,brittain09,perdew09} Current functionals can successfully predict semiconductor band gaps of inorganic crystals to within 0.1 eV, but much of this accuracy can be attributed to a fortuitous `cancellation of errors'.\cite{ku02}
In the context of organometallic complexes, one of the worrying consequences of these problems is that (TD)DFT tends to overestimate the delocalisation of electron orbitals.\cite{cohen08,perdew09} 
In order to better understand both the attributes and the limitations of (TD)DFT calculations we finish this section by briefly reviewing three calculations \cite{hay02,obara06,lo06} - highlighting on both the insights gained and the limitations of the calculations.

Hay \cite{hay02} used TDDFT to investigate the effects of perturbing an Ir(ppy)$_3$ complex by changing one ligand, replacing one phenylpyridine (ppy) with either acetoylacetonate (acac) or benzyolacetonate (bza). He found that the HOMOs of each of these complexes have around 50\% metal character, the rest being ligand $\pi$, while the LUMOs were predicted to be purely ligand $\pi^*$ orbitals. Thus, he argued that excitations in these complexes are hybrid MLCT-LC. When one of the ppy ligands is replaced by acac or bza Hay predicted only small changes in the HOMO energy. However the HOMO-1 and HOMO-2, which are degenerate in Ir(ppy)$_3$ due to its $C_{3}$ symmetry, are split and stabilized by the symmetry perturbation introduced by the new ligand. In Ir(ppy)$_3$ the LUMO was predicted to be a combination of all three ligands, as were the LUMO+1 and LUMO+2: a degenerate $E_g$ pair. In Ir(ppy)$_2$(acac) the LUMO and LUMO+1 were predicted to be degenerate and localised on the two remaining ppy ligands. This degeneracy is rather surprising as this molecule has $C_2$ symmetry and thus no symmetry required degenerices. 
In contrast the LUMO+2 was predicted to be localised on the acac ligand. Hay found that in Ir(ppy)$_2$(bza) the LUMO is localised on the bza ligand. He argued that this localisation of the LUMO in Ir(ppy)$_2$(bza) is the reason that this complex has a much lower PLQY than Ir(ppy)$_2$(acac) or Ir(ppy)$_3$ ($<0.01$ in Ir(ppy)$_2$(bza),\cite{lamansky01} compared to 0.3 and $>0.4$ in Ir(ppy)$_2$(acac)\cite{lamansky01} and Ir(ppy)$_3$\cite{king85}, respectively). However, it is not known whether this change comes via a supression of the radiative decay rate or an enhancement of the non-radiative decay rate.

Obara \emph{et al}. \cite{obara06} investigated a series of iridium complexes with both bi- and tri-dentate ligands. They used TDDFT calculations to investigate the large transition dipole moment from states that are predominantly triplet found in many organometallic complexes. Following previous discussions of rhodium complexes,\cite{komada86,miki93} they considered two mechanisms by which spin-orbit coupling might allow triplet radiative emission. One is direct coupling between a LC triplet and MLCT singlet. The other is the coupling between the MLCT component of a predominantly LC triplet and the MLCT singlet. They then found that the direct LC to MLCT spin-orbit coupling is unlikely to be large enough to explain the bright triplets observed experimentally. Obara \emph{et al}. conclude that a MLCT triplet to MLCT singlet coupling is the most reasonable explanation for the intense phosphorescence of the complexes they investigated. These results support the hypothesis of Finkenzeller \emph{et al}. that increasing the MLCT character of the triplet will increase its radiative decay rate and hence its potential as an emitter material.\cite{obara06}

Lo \textit{et al}. \cite{lo06} investigated a series of iridium complexes with aryltriazolyl-based ligands experimentally and with DFT calculations. They found, experimentally, that  small chemical changes cause large variations in both the radiative decay rate (discussed further in the following section) and the non-radiative decay rate. 
 They suggested  a correlation between the measured photoluminescent quantum yield (PLQY) and the calculated molecular orbital energy gaps.\cite{lo06} However, they found that this correlation only remains as long as the non-radiative decay rate does not vary much. In addition, they found that the HOMO of Ir(ppy)$_3$ has five times as much electron density on the phenyl ring as on the pyridyl ring, while the LUMO has three times as much electron density on the pyridyl ring as on the phenyl. Thus, substituents that stabilize the phenyl ring will lower the energy of the HOMO more than that of the LUMO. Similarly, destabilizing the pyridyl ring raises the energy of the LUMO more than that of the HOMO. Therefore both stabilizing the phenyl ring and destabilizing the pyridyl ring result in a blue-shift of the spectrum. Thus, they were able to explain why one can tune the HOMO and LUMO of Ir(ppy)$_3$ somewhat independently.

\section{Model Effective Hamiltonians}

Another approach to modeling the optoelectronic properties of organometallic complexes is to construct a model with fewer states but an accurate treatment of the electronic correlations. This contrasts with first principles calculations, which include several basis states for each atom but neglects some electronic correlations. The small number of degrees of freedom in such semi-empirical models allows one to make fewer approximations on the interactions and correlations in the model. It also allows one to identify key trends that describe broad classes of materials. This approach has proven itself incredibly powerful in wide areas of materials science. For example, the Anderson single impurity model can describe a wide range of systems including magnetic impurities in metals, quantum dots in semiconductor heterostructures, carbon nanotubes, and single molecule transistors.\cite{hewsonbook,kouwenhoven01}

In principal an effective model Hamiltonian is found by starting with the exact Hamiltonian and `integrating out' high energy states.\cite{powell09book} This procedure is computationally expensive,\cite{freed83} so often one simply chooses a reduced basis set, motivated by the physical processes one wishes to capture.\cite{powell09book}
DFT can be used to estimate the values (or trends in values) of some of the parameters of these effective models. The model Hamiltonian can then solved, retaining correlations that the approximate DFT functional does not include.\cite{scriven09,gunnarssonbook,brocks04,canocortes07,scriven09B}

Identifying the frontier orbitals which dominate the photophysics is one of the most significant steps of the effective model approach.
In this reduced basis set one can define an effective Hamiltonian with just a few parameters. Conjugated polyenes have been investigated in this way via the H\"uckel, Hubbard, Heisenberg and Pariser-Parr-Pople models.\cite{powell09book} This approach has been applied to organometallic complexes, for example mixed valence binuclear systems including magnetic atoms in proteins (Hubbard and double exchange models)(Ref. \onlinecite{blondin90}), molecular magnets (Ref. \onlinecite{nagao00}), Anderson impurity models for cobalt based valence tautomers (Ref. \onlinecite{labute02}), and a series of metal-cored bipyridine complexes (Ref. \onlinecite{kober82}). It has also been shown recently that this approach naturally explains the sensitivity of the photophysical properties of organometallic complexes to small chemical changes.\cite{jacko10a}

To correctly describe the character of the excited states the model must capture the key interactions. There are many important features of the system that might be included in such a model, for example electronic `hopping' terms between the frontier molecular orbitals, direct Coloumb interactions between electrons in those orbitals, spin interactions, and relativistic effects such as spin-orbit coupling. The relative energy scales of these various interactions will define the composition of the excited states and therefore their properties. Again, some examples will help to illustrate both the power and the limitations of this approach.

Kober and Meyer \cite{kober82} studied a simple model Hamiltonian for $\mathcal{D}_3$ symmetric bipyridine (bpy) complexes.
Their basis consists of 3 HOMOs and 3 LUMOs (two of E symmetry and one of A). The HOMOs are of mixed metal-ligand character, while the LUMOs are pure ligand $\pi^*$. The parameters of the model are the HOMO-LUMO gap, the splitting of the A and E HOMOs, the splitting of the A and E LUMOs, the singlet-triplet splitting and the degree of spin-orbit coupling.
To make progress this model has to be fitted experimental spectra. This is a significant limitation. However, the fit reproduces the experimental data quite well. %, as shown in Fig. \ref{fig:kober82_modelfit_Rubpy}.
 On the basis of this model, they concluded that although there is appreciable singlet-triplet mixing, labelling states as `singlet' or `triplet' is still reasonable, and can be helpful in understanding the photophysics. They also found that although the initial excitation occurs to a symmetric combination of bpy ligands, the relaxed excited state does not necessarily have this symmetry.
%
%\begin{figure}
%	\centering
%		\includegraphics[width=6.5cm]{kober82_modelfit_Rubpy}
%	\caption{Comparison of polarized experimental spectra of Ru(bpy)$_3$$^{2+}$ with those calculated from a effective model Hamiltonian, as in Ref. \onlinecite{kober82}. Their straightforward four parameter model Hamiltonian fits the experimental data well.}
%	\label{fig:kober82_modelfit_Rubpy}
%\end{figure}

\begin{figure}
	\centering
		\includegraphics[width=6.5cm]{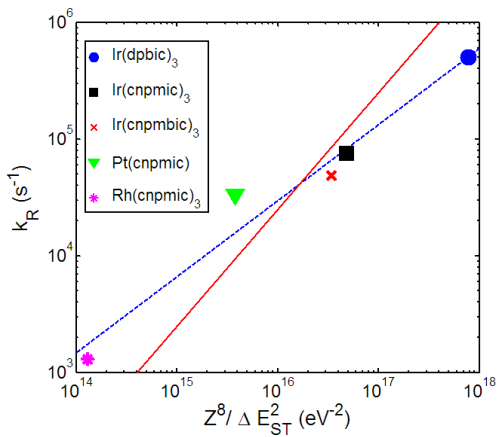}
	\caption{Experimental phosphorescent radiative rates versus calculated singlet-triplet gaps scaled by atomic number Z, as in Ref. \onlinecite{haneder08}, whose data we show here. Ref. \onlinecite{haneder08} discussed this correlation for the iridium and platinum complexes, all of which have similar values of Z. One can easily extend this to complexes with very different values of Z (and therefore spin-orbit coupling) by including the $Z^4$ dependence in the transition dipole moment.\cite{yersinbook} This shows that reductions in the singlet-triplet gap are correlated with increases in the radiative decay rate. The solid red line is a fit of the data to the predicted $k_R \propto Z^8 / \Delta E_{ST}^2$ behaviour. The dashed blue line is a line of best fit to $k_R \propto (Z^8 / \Delta E_{ST}^2)^{\alpha}$ with $\alpha=2/3$, indicating that although that, over this data set, the predictions are only followed qualitatively.}
	\label{fig:haneder08_gapvrate}
\end{figure}

Haneder \emph{et al}. \cite{haneder08} investigated the correlation between the S$_1$-T$_1$ gap, $\Delta E_{ST}$, and the radiative decay rate of the triplet, $k_R$. They considered a three level system consisting of the S$_0$, S$_1$ and T$_1$ states and calculated the radiative decay rate of the triplet state by including the spin orbit coupling perturbatively. They found (within their small sample of molecules) that there was indeed a qualitative trend similar to their prediction that $k_R\propto\Delta E_{ST}^2$, cf. Fig. \ref{fig:haneder08_gapvrate}.  The strength of the spin-orbit coupling varies with the atomic number, $Z$, of the transition metal as $Z^4$, while the radiative decay rate is proportional to the square of the transition dipole moment. Therefore, one should also expect that $k_R\propto Z^8$, Fig. \ref{fig:haneder08_gapvrate} shows that this is followed as to the same accuracy Haneder \emph{et al}.'s original prediction is. Thus, this very simple model leads to two straightforward design principles: (i) minimize the singlet-triplet gap to maximize the triplet radiative rate and (ii) maximise the atomic number of the transition metal.

%| - = + = - = + = - = + = - = + = - = + = - = + = - = + = - |
%|                                                           |
%| 							    Our stuff starts here									   |
%|                                                           |
%| - = + = - = + = - = + = - = + = - = + = - = + = - = + = - |

We have recently applied a  semi-empirical approach to modelling the optoelectronic properties of organometallic complexes.\cite{jacko10a} We employed a basis of effective fragment orbitals, one pair of frontier ligand orbitals and one metal orbital. The Hamiltonian for this model explicitly included the hopping between the ligand orbitals and the metal, an effective spin exchange interaction between the ligands HOMO and LUMO, and the renormalised direct Coloumb interactions on and between every site. It is important to realise that these effective renormalised parameters will in general be very different from the bare parameters used in first principles methods \cite{powell09book,scriven09}. Differences between complexes are encapsulated into the effective parameters of this three orbital model. Nevertheless, this model Hamiltonian can be solved exactly. 

\begin{figure}
	\centering
		\includegraphics[width=6.65 cm]{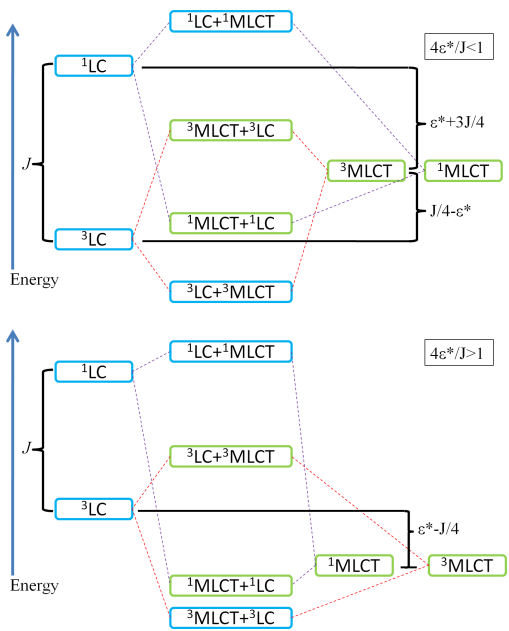}
	\caption{Correlation diagram, showing triplet basis states on either side and the resulting hybrid orbitals in the center. The upper figure corresponds to the situation $\varepsilon^*/J < 1/4$, where the $T_1$ state is predominantly LC, while the lower figure shows $\varepsilon^*/J > 1/4$ where the $T_1$ state is predominantly MLCT. In the intermediate case $\varepsilon^* /J \simeq 1/4$, both the $T_1$ and $T_2$ triplet states are strongly mixed MLCT-LC hybrids.}
	\label{fig:modeldiagrams_11_hybrids}
\end{figure}

%We find that within the singlet spectra this conclusion may indeed be realistic. However, in the triplet states we find that the character of the states can be strongly mixed. [Fig 5: MLCTness] 
For the parameters characteristic of the organometallic complexes used in OPV and OLED devices, the lowest excited singlet eigenstates are pure MLCT and LC, while the triplet states are typically strongly mixed. The parameter that controls the character of the triplet states is the $^3 MLCT$-$^3 LC$ energy gap (as shown in Fig. \ref{fig:modeldiagrams_11_hybrids})
\beq \label{eq:epstar}
\varepsilon^* - J/4
\eeq
where $\varepsilon^*$ is the renormalized effective ligand HOMO-metal energy gap and $J$ is the exchange interaction between an electron in the ligand HOMO and an electron in the ligand LUMO. $J$ determines the gap between the lowest singlet triplet excited states in the isolated complex. The $^3 MLCT$ and $^3 LC$ states are not eigenstates. Thus the size and sign of the gap between them determines the character of the lowest excited triplet state, $T_1$. Note that a molecular orbital theory would predict that singlets and triplets are degenerate ($J=0$), a prediction that is not borne out experimentally.

%When $\varepsilon^*/J = 1/4$ the $^3 MLCT$ and $^3 LC$ sates are degenerate. Thus, for $J\simeq1/4$ the lowest triplet state $T_1$ is an equal hybrid of LC and MLCT. For $\varepsilon^*/J >1/4$, $T_1$ is predominantly MLCT, whereas for $\varepsilon^*/J < 1/4$ it is predominantly LC (illustrated in Fig. \ref{fig:modeldiagrams_11_hybrids}).\cite{jacko10a} Most of the parameters of our model can be extracted from either basic theory or from comparison to experiments on the isolated ligand. The only unknown parameter is $\varepsilon^*$, which is explicitly a property of the complex. However, it may be possible to estimate the variation in $\varepsilon^*$ by a comparison with the Hammett constants of the substituents.\cite{dahne07}

%Moreover, the degree of this mixing is strongly dependent on parameters which cannot be measured in fragments of the complex. [Fig 6: Correlation diagrams, 2 regimes] 
Most of the parameters of our model can be extracted from either basic theory or from comparison to experiments on the isolated ligand \cite{jacko10a}.
However, the parameter $\varepsilon^*$ cannot be predicted from studies of fragments of the complex, it is a property of the complex, the particular combination of metal and ligand. In particular it depends on the complex-specific properties the HOMO-metal energy gap $\varepsilon$ and the strength of the LUMO-metal Coloumb interactions. \fig{modeldiagrams_11_hybrids} shows the two important parameter regimes of the model. In one ($\varepsilon^*/J<1/4$) the lowest excited triplet is predominantly LC, while in the other ($\varepsilon^*/J>1/4$) the lowest excited triplet is MLCT. In the latter case, the energy gap between the lowest excited triplet and lowest excited singlet becomes small.

%%%%%%%%read to here

\begin{figure}
	\centering
		\includegraphics[width=6.5cm]{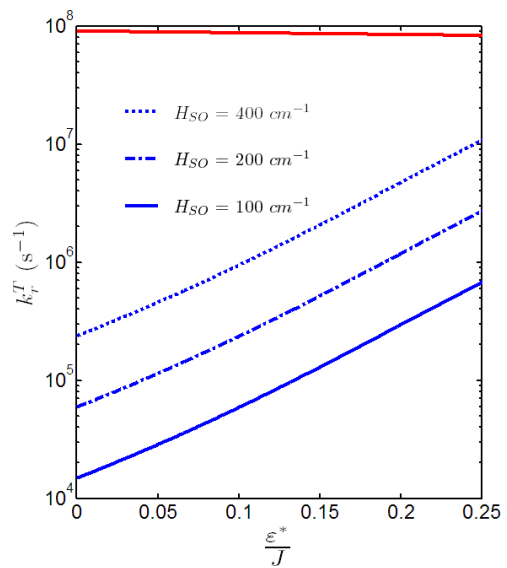}
	\caption{Triplet radiative rate $k_r^T$ (lower curves, blue) and singlet radiative rate (upper curve, red) as a function of $\varepsilon^*/J$ for various values of $H_{SO}$ solved to first order in perturbation theory. We have chosen $ \left(\expect{\hat{r}}_{^1 MLCT^1} - \expect{\hat{r}}_0\right) \simeq 20$ \AA $ $ to reproduce a singlet radiative rate of $\sim 10^*$ s$^{-1}$. The triplet radiative rate increases exponentially as $\varepsilon^*/J$ increases, up until the point where the lowest singlet and triplet are nearly degenerate $\varepsilon^*/J \sim 0.25$ (\ie both MLCT) at which point the perturbative solution becomes invalid. As the strength of the spin-orbit coupling increases, the triplet radiative rate rapidly increases. Around $\varepsilon^*/J \sim 1/4$ perturbation theory breaks down, and a more complete treatment is required. }
	\label{fig:tripletlifetime}
\end{figure}

This variation in triplet state character can have profound effects on the optoelectronic properties of the complex. $\varepsilon^*/J$ can have a dramatic effect on the triplet's radiative decay rate, as illustrated in Fig. \ref{fig:tripletlifetime}.\cite{jacko10a}    
As $\varepsilon^*/J$ increases, the lowest triplet state's MLCT character increases and it gets closer in energy to the MLCT singlet state, both of which serve to increase the effects of spin-orbit coupling. In the regime in which the spin orbit coupling can be included perturbatively ($\varepsilon^*/J \lesssim 1/4$), the triplet radiative rate increases by more than an order of magnitude when $\varepsilon^*/J$ is varied from 0.1 to 0.25. In this regime the radiative rate can be shown to vary with the singlet-triplet gap to the fourth power, which, on the scale in the figure, appears linear on a semilogarithmic plot.\cite{jacko10a}
In the regime $\varepsilon^*/J < 1/4$, the degree of MLCT character in the triplet is perturbative, and we find agreement with the hypothesis of Finkenzeller \emph{et al}.: increasing the MLCT character increases the radiative rate of the emitting triplet.\cite{finkenzeller07}

\begin{figure}
	\centering
		\includegraphics[width=6.5cm]{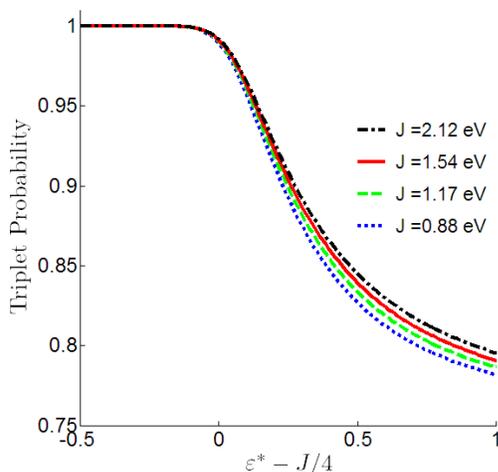}
	\caption{Triplet probabilities versus the $^3 LC - ^3 MLCT$ energy gap $\varepsilon^* - J/4$ for various values of $J$, corresponding to the ligands ppy ($J = 2.12$ eV), thpy ($J = 1.54$ eV), fluorene ($J = 1.17$ eV) and bzq ($J = 0.88$ eV), all at 300K.}
	\label{fig:tripletprob_VSepsilonstar}
\end{figure}

We also determined the effect of the parameter $\varepsilon^*/J$ on triplet production probability following charge injection. The characteristic time for singlet-triplet intersystem crossing is very short compared to the singlet and triplet lifetimes, so there will be a Boltzmann distribution of excited states. The lowest singlet and triplet are well separated (compared to $k_B T$) from other excited states, so one only needs to consider these two lowest excited states. When the singlet-triplet gap is much less than $k_B T$ we have the expected 75\% triplet probability to 25\% singlet probability. On the other hand, if the singlet-triplet gap is much larger than $k_B T$, we have nearly 100\% triplet probability. This means that injected charges will almost always decay to the ground state via the lowest triplet state if the singlet-triplet gap is larger than $k_B T$. In our model a large singlet-triplet gap corresponds to the regime $\varepsilon^*/J < 1/4$, as shown in Fig. \ref{fig:tripletprob_VSepsilonstar}.

Therefore we have seen that this model gives a firm mathematical basis to some of the key ideas in the literature. The hypothesis of Finkenzeller \emph{et al}. that the triplet is predominantly LC with a perturbative MLCT contribution is reproduced in the $\varepsilon^*/J < 0.25$ regime of the model. In this regime, increasing $\varepsilon^*/J$ simultaneously decreases the singlet-triplet gap and increases the triplets MLCT character. Both of these lead to an increased triplet radiative rate, in agreement with the hypotheses in Refs. \onlinecite{finkenzeller07} and \onlinecite{haneder08}. Note that increasing $\varepsilon^*/J$ does not guarantee an increase in PLQY, as we did not attempt to predict changes in the non-radiative decay rate (as this requires a knowledge of the non-radiative decay path on the excited state potential energy surface). 
Furthermore, one should expect large changes in the non-radiative rate to be caused by small changes in the chemistry as non-radiative decay is an activated process and so will depend exponentially on energy gaps. 

\section{Conclusions}

First principles electronic structure and effective model Hamiltonians provide complementary methods to understand the optical properties of organometallic complexes with applications in OPV and OLED technologies. (TD)DFT calculations have for a long time been the workhorse in the field and have led to important insights, such as, those discussed above concerning how changes to the symmetry or chemistry of a complex can effect its PLQY. However, semi-empirical models can also provide significant insights: particularly for understanding trends across broad classes of materials. Here we have highlighted just a few examples: The radiative decay rate is sensitive to small changes in the ligand chemistry because small chemical changes drive large changes in the degree of MLCT character in the triplet state. The same physics determines the fraction of injected charges that form triplet excitations.

%This paper reviews recent models of organometallic complexes for optoelectronic applications and some of the insights gained from these models. Effective model Hamiltonians can provide qualitative trends in the optoelectronic properties, and explain the sensitivity of the excited state properties to small chemical substitutions. We have discussed the effect of mixing MLCT and LC triplet states on the radiative lifetime. We also discussed the change in the triplet production probability upon charge injection in the presence of singlet-triplet splitting, away from the oft-quoted 75\% proportion.

\begin{acknowledgments}
We are grateful to Paul Burn, Hamish Cavaye, Ben Langley, Lawrence Lo, A. Bernardus Mostert, Seth Olsen, Paul Schwenn, Paul Shaw and Arthur Smith for helpful discussions. R. H. M. was the recipient of an Australian Research Council (ARC) Australian Professorial Fellowship (project no. DP0877875). 
B. J. P. was the recipient of an ARC Queen Elizabeth II Fellowship (project no. DP0878523). 
\end{acknowledgments}

\bibliographystyle{rsc}
\bibliography{refs}

\end{document}